\documentclass[prb,clbiba,12pt,psfig]{article}
\usepackage{epsfig}
\usepackage{epstopdf}
\usepackage{graphicx}
\usepackage{float}
\usepackage{subcaption}
\usepackage{overcite}
\begin{document}
\large
\title{\bf Numerical Prediction of NaCl Solubility in Water Using the  TIP4P/05 Water Model NaCl/Madrid.}
\maketitle{
{{\noindent Ra\'ul Fuentes-Azcatl \\
{\it }\\

{\footnotesize Universidad Aut\'onoma Metropolitana, campus Xochimilco, Calz. del Hueso 1100, Coapa, Villa Quietud, Coyoac\'an, 04960 Ciudad de M\'exico, CDMX\\}
\maketitle


\vspace{0.5in} \noindent \\ Electronic mail: rfuentes@correo.xoc.uam.mx

\newpage

\begin{abstract}

The solid--liquid equilibrium and solubility of NaCl in water were investigated using molecular dynamics simulations with the NaCl/Madrid force field combined with the TIP4P/05 water model. Particular attention was given to the dependence of the calculated solubility on the intermolecular cut-off radius, $r_{\mathrm{cut}}$, used to identify the ions belonging to the crystalline phase. The simulations were performed starting from a NaCl crystal immersed in an aqueous NaCl solution, following the configuration proposed by Manzanilla-Granados et al. The system was initially equilibrated and subsequently analysed over an extended simulation trajectory. To establish a physically consistent criterion for distinguishing the solid and aqueous phases, the structural information obtained from the radial distribution function, $g(r)$, was incorporated into the analysis. The first minimum following the principal Na--Cl correlation peak was used to characterise the spatial extent of the ionic coordination environment and to constrain the range of physically meaningful $r_{\mathrm{cut}}$ values. The number of ions remaining in the crystalline phase was then evaluated as a function of $r_{\mathrm{cut}}$, allowing the corresponding equilibrium solubility to be determined. The resulting dependence exhibits a well-defined inflection point, which provides a criterion for identifying the transition between predominantly crystalline and dissolved ionic populations. This approach yields an estimate of the NaCl solubility directly from the structural and compositional evolution of the solid--liquid system and provides a physically motivated alternative to selecting the cut-off distance arbitrarily. The results demonstrate that the combination of radial distribution functions and the $r_{\mathrm{cut}}$-dependent analysis provides a useful framework for characterising the solid--liquid behaviour and determining the solubility of ionic crystals from molecular dynamics simulations.

\end{abstract}

\section{Introduction}

The solubility of inorganic salts in water is a fundamental thermodynamic property that plays an important role in chemical, environmental, geological, and industrial processes. Among aqueous electrolyte systems, sodium chloride in water represents one of the most extensively studied model systems for understanding salt dissolution, crystallization, and solid--liquid equilibrium. Despite the apparent simplicity of the NaCl--H$_2$O system, an accurate prediction of its solubility from molecular simulations remains challenging because it requires a consistent description of both the crystalline salt and the concentrated aqueous electrolyte solution.

\begin{equation}
a_{\pm}
=
\gamma_{\pm}
\frac{m}{m^{\circ}},
\label{eq:mean_activity}
\end{equation}
where $\gamma_{\pm}$ is the mean ionic activity coefficient, $m$ is the NaCl molality, and $m^{\circ}$ is the standard molality. Consequently, the solubility reflects the balance between the thermodynamic stability of the ionic crystal and the stabilization of the dissolved ions in water \cite{Rogers2005,Rogers2011,Palmer2016}.

The thermodynamic description of electrolyte solutions becomes increasingly complex at high salt concentrations because the ions cannot be treated as independent species. Long-range electrostatic interactions, ion--ion correlations, hydration effects, and deviations from ideality contribute significantly to the chemical potential of the dissolved salt \cite{Rogers2005,Moucka2011,Rogers2011,Palmer2016}. Consequently, an accurate molecular model must simultaneously provide a realistic description of the ionic crystal, the hydration structure of the dissolved ions, and the collective properties of the concentrated electrolyte solution.

Molecular dynamics (MD) simulations provide an effective framework for investigating this balance at the molecular level. Unlike macroscopic thermodynamic descriptions, MD simulations allow the structural and dynamical properties of the crystalline and aqueous phases to be examined explicitly. In particular, ion--ion correlations, ion--water interactions, local coordination environments, and structural changes at the solid--liquid interface can be directly characterized. These microscopic properties are closely related to the relative stability of the two phases and therefore provide valuable information for understanding the molecular mechanisms governing salt dissolution and crystallization.

Several molecular simulation approaches have been developed to determine the solubility of aqueous electrolytes. Free-energy calculations, thermodynamic integration, expanded-ensemble methods, and osmotic-ensemble approaches have been employed to determine chemical potentials and phase equilibria in electrolyte solutions \cite{Rogers2005,Moucka2011,Rogers2011,Palmer2016}. These methods provide a rigorous thermodynamic framework for determining solubility, but they generally require the calculation of free-energy differences or chemical potentials associated with the solid and solution phases. Such calculations can become computationally demanding, particularly for concentrated electrolyte solutions and crystalline systems, where long equilibration times and adequate sampling of both phases are required.

An alternative strategy is to determine the solid--liquid equilibrium directly through molecular simulation. In the direct coexistence approach, an explicit crystalline phase is placed in contact with an aqueous electrolyte solution within the same simulation cell. The resulting solid--liquid interface provides a direct representation of the competition between dissolution and crystallization. When the solution is undersaturated, dissolution of the crystal is thermodynamically favored and ions are transferred from the solid to the aqueous phase. Conversely, when the solution is supersaturated, incorporation of dissolved ions into the crystalline phase is favored, resulting in crystal growth. At the saturation condition, the rates of dissolution and crystallization balance on average, and the crystalline phase remains stable over the simulation time scale \cite{Ladd1989,Davidchack2003,Davidchack2005,ManzanillaGranados2015,Espinosa2016}.

The direct coexistence method has been successfully applied to the calculation of the solubility of NaCl in water. Manzanilla-Granados et al.~\cite{ManzanillaGranados2015} investigated the direct coexistence of crystalline NaCl and aqueous solutions and demonstrated that the equilibrium solubility can be obtained by monitoring the evolution of the crystalline phase at different solution compositions. More recently, Espinosa et al.~\cite{Espinosa2016} systematically investigated the calculation of solubilities using direct coexistence simulations and emphasized the importance of the initial configuration, equilibration, and structural characterization of the solid and liquid phases. These studies demonstrate that direct coexistence provides a computationally convenient alternative to explicit free-energy calculations while retaining a direct connection with the underlying phase equilibrium.

An important aspect of direct coexistence simulations is the identification and characterization of the crystalline phase during the simulation. Because dissolution and crystallization occur continuously at the solid--liquid interface, the number of ions assigned to the crystal can fluctuate substantially with time. Consequently, the structural criterion used to distinguish crystalline ions from liquid-like ions can influence the apparent evolution of the solid phase and, ultimately, the determination of the solubility. The definition of the solid phase is therefore not merely a post-processing detail but an important component of the coexistence analysis.

Different structural criteria have been proposed to identify crystalline particles in molecular simulations. Local bond-order parameters based on the Steinhardt formalism provide a measure of the orientational order surrounding individual particles and have been widely used to distinguish solid-like and liquid-like environments \cite{Steinhardt1983,tenWolde1995}. In the context of aqueous NaCl, Espinosa et al.~\cite{Espinosa2016} employed a local order parameter based on $q_4$ to identify ions belonging to the crystalline phase. An ion was classified as solid-like when it exhibited at least four neighboring ions satisfying
\begin{equation}
q_4(i)\cdot q_4^{*}(j)>0.35.
\label{eq}
\end{equation}
This criterion provides a local structural measure of crystallinity and, importantly, does not rely exclusively on a prescribed radial distance between ions.

A complementary approach is to identify the crystalline network using a geometrical connectivity criterion based on the radial distribution function. For an ionic crystal such as NaCl, the Na--Cl radial distribution function, $g_{\mathrm{Na-Cl}}(r)$, provides direct information about the characteristic distances between oppositely charged ions and about the local coordination structure of the crystal. Radial distribution functions have long been used to characterize the microscopic structure of ionic solutions and crystalline phases \cite{Allen2017,Leach2001}. A cutoff distance can therefore be introduced according to
\begin{equation}
r_{\mathrm{Na-Cl}} < r_{\mathrm{cut}},
\label{eq}
\end{equation}
where ions satisfying this condition are considered structurally connected through the Na--Cl network.

The choice of $r_{\mathrm{cut}}$ is consequently an important methodological parameter. If the cutoff is too small, ions that are structurally associated with the crystalline phase may be excluded from the solid network. Conversely, an excessively large cutoff can connect liquid-like ions to the crystal and artificially increase the apparent crystalline population. This issue is particularly relevant for NaCl because the radial distribution function contains distinct structural features associated with short-range coordination and longer-range ionic ordering. Therefore, the use of a fixed cutoff without examining its structural basis may introduce a systematic bias into the determination of the crystalline population.

In the present work, the cutoff distance is therefore not assumed a priori. Instead, the Na--Cl radial distribution function is first analyzed to identify the characteristic structural distances of the system, and a systematic range of $r_{\mathrm{cut}}$ values is subsequently investigated. The effect of this parameter on the number of ions assigned to the crystalline phase and on the observed evolution of the crystal is explicitly evaluated. In this way, the sensitivity of the calculated solubility to the structural definition of the solid phase can be quantified.

This distance-based criterion is complementary to the local bond-order approach used by Espinosa et al.~\cite{Espinosa2016}. The two approaches characterize different aspects of crystallinity: the $q_4$ criterion identifies ions according to their local orientational order, whereas the present approach identifies the crystalline phase through the geometrical connectivity of the Na--Cl network. The comparison is particularly relevant because the two definitions need not assign exactly the same set of ions to the crystalline phase, especially near the solid--liquid interface where structural order changes continuously.

The structural definition of the crystalline phase is especially important when the objective is to determine solubility from the temporal evolution of a finite crystal. In a direct coexistence simulation, the equilibrium condition is not obtained from a single instantaneous configuration but from the long-time behavior of the crystal. Thus, the identification of crystalline ions must be sufficiently robust to distinguish genuine dissolution or growth from fluctuations associated with interfacial rearrangements. A systematic analysis of $r_{\mathrm{cut}}$ provides a means of evaluating this sensitivity and of establishing a reproducible criterion for subsequent coexistence calculations.

The direct coexistence approach is particularly useful for assessing the consistency of molecular interaction models because the same force field simultaneously describes the crystalline and aqueous phases. The calculated solubility consequently reflects the relative stability of these two environments and provides a stringent test of the balance among ion--ion, ion--water, and water--water interactions. This is especially relevant for concentrated NaCl solutions, where strong ionic correlations and deviations from ideal solution behavior become increasingly important. A force field that accurately reproduces the structural properties of the NaCl crystal or the properties of dilute aqueous solutions does not necessarily guarantee an accurate prediction of the saturation concentration.

The NaCl--H$_2$O system therefore provides an appropriate model for evaluating the transferability and internal consistency of molecular interaction potentials across different thermodynamic environments. The NaCl crystal is characterized by a highly ordered ionic lattice, whereas the aqueous phase contains hydrated ions whose local structure and interactions vary strongly with salt concentration. The solubility represents the macroscopic manifestation of the competition between these two environments and consequently provides information that cannot be obtained from the properties of either phase independently.

In the present work, the solubility of NaCl in water is investigated using a direct solid--liquid coexistence approach based on molecular dynamics simulations. An explicit NaCl crystal is placed in contact with an aqueous NaCl solution, and the stability and evolution of the crystalline phase are monitored for different salt concentrations. Particular attention is given to the structural criterion used to identify ions belonging to the crystalline phase. The Na--Cl radial distribution function is employed to determine characteristic structural distances, while a systematic variation of $r_{\mathrm{cut}}$ is used to evaluate its influence on the identification of the crystalline phase and on the calculated solubility.

The resulting dependence of the calculated solubility on $r_{\mathrm{cut}}$ is used to establish a physically motivated criterion for identifying the crystalline phase. This analysis makes it possible to distinguish the structural definition of the crystal from the thermodynamic determination of the saturation point and to assess explicitly the uncertainty associated with the classification of ions at the solid--liquid interface. The combination of structural and thermodynamic analyses provides a more robust framework for determining the NaCl solubility and for evaluating the molecular model used to describe the NaCl--H$_2$O system.

\section{  force Fields}

\subsection{NaCl , force fields}

The force field used in this work to represent NaCl in aqueous solution consists of a set of pairwise interaction potentials defined between the corresponding interaction sites. In classical molecular dynamics simulations, these interactions are commonly described by a combination of Lennard--Jones (LJ) and Coulombic terms, which account for short-range van der Waals interactions and long-range electrostatic interactions, respectively \cite{Leach2001,Allen2017}.

For the aqueous phase, the NaCl/Madrid model developed by Benavides et al. is used in combination with the TIP4P/05 water model \cite{Benavides2017,Abascal2005}. The Madrid model was specifically developed for aqueous NaCl solutions using TIP4P/05 water and employs scaled ionic charges of $\pm 0.85e$ to account effectively for electronic polarisation effects that are not explicitly represented in a non-polarisable force field \cite{Benavides2017}. The model was assessed against a broad range of experimental properties of aqueous NaCl solutions, including the equation of state, density, enthalpy of solution, activity coefficients, radial distribution functions, solubility, surface tension, self-diffusion coefficients, and viscosity. In the present work, the NaCl/Madrid--TIP4P/05 model is evaluated with respect to the density, ionic structure, and self-diffusion properties of aqueous NaCl solutions.

The pairwise interaction potential is given by

\begin{equation}
u(r_{ij}) =
4\varepsilon_{ij}
\left[
\left(\frac{\sigma_{ij}}{r_{ij}}\right)^{12}
-
\left(\frac{\sigma_{ij}}{r_{ij}}\right)^6
\right]
+
\lambda_i\lambda_j
\frac{q_iq_j}{4\pi\varepsilon_0 r_{ij}},
\label{eq:potential}
\end{equation}

where $r_{ij}$ is the distance between interaction sites $i$ and $j$, $q_i$ and $q_j$ are the corresponding electric charges, $\varepsilon_0$ is the vacuum permittivity, $\varepsilon_{ij}$ is the Lennard--Jones (LJ) energy parameter, and $\sigma_{ij}$ is the LJ size parameter. The factors $\lambda_i$ and $\lambda_j$ account for the scaling of ionic charges, when applicable. Charge scaling for monovalent ions has been employed in molecular dynamics simulations to account effectively for electronic polarisation effects that are not explicitly represented in conventional non-polarisable force fields \cite{nacle,kbre,HgCl2e}.

The use of scaled ionic charges in non-polarizable force fields is related to the electronic continuum correction (ECC) and molecular dynamics in an electronic continuum (MDEC) approaches introduced by Leontyev and Stuchebrukhov \cite{LeontyevStuchebrukhov2009,LeontyevStuchebrukhov2010}. These approaches provide an effective way of accounting for electronic polarization effects that are not explicitly represented in conventional fixed-charge force fields. In this framework, the ionic charges are reduced from their formal values to incorporate, in an effective manner, the electronic screening of the condensed phase. The NaCl/Madrid model adopts this type of charge scaling, with the ionic charges modified by a common scaling factor $\lambda_C$.

The interaction potentials for water and ions are assumed to be mutually compatible, meaning that they share the same functional form. Consequently, the cross interactions between water molecules and ions are determined using the Lorentz--Berthelot (LB) combining rules for the conformal Lennard--Jones potential \cite{Leach2001,Allen2017}:

\begin{equation}
\sigma_{\alpha\beta}=
\frac{\sigma_{\alpha\alpha}+\sigma_{\beta\beta}}{2},
\label{eq:LB_sigma}
\end{equation}

and

\begin{equation}
\varepsilon_{\alpha\beta}=
\sqrt{\varepsilon_{\alpha\alpha}\varepsilon_{\beta\beta}}.
\label{eq:LB_epsilon}
\end{equation}

Here, $\alpha$ and $\beta$ denote two different interaction sites, such as an ion and an atom belonging to a water molecule. These combining rules allow the cross Lennard--Jones parameters to be obtained directly from the corresponding parameters of the individual species, without introducing additional independent parameters for each ion--water pair. This approach is consistent with the conventional treatment of nonbonded interactions in classical molecular mechanics force fields \cite{Leach2001,Allen2017}.

The NaCl salt is represented using a rigid, non-polarizable ionic model, in which each spherical ion is described by a single interaction site located at its center. The cation and anion carry charges of $q_i=\pm 1e$, where $e$ denotes the elementary charge. For the NaCl/Madrid parametrization, the absence of explicit electronic polarization is accounted for by scaling the electrostatic interaction through a screening factor, $\lambda_i=\lambda_C$, applied equally to the sodium and chloride ions. Consequently, the NaCl/Madrid model is characterized by three main parameters: the charge-scaling factor $\lambda_C$, the Lennard--Jones size parameter $\sigma_{\mathrm{LJ}}$, and the Lennard--Jones energy parameter $\varepsilon_{\mathrm{LJ}}$.

The parameters used for the NaCl/Madrid force field are listed in Table~\ref{tab:nacl_parameters}.

\begin{table}[ht]
\centering
\caption{Force-field parameters of the NaCl/Madrid model.}
\label{tab:nacl_parameters}
\begin{tabular}{|lcc|}
\hline
\multicolumn{3}{|c|}{NaCl/Madrid} \\
\hline
Parameter & Na$^+$ & Cl$^-$ \\
\hline
$q/e$ & $+1$ & $-1$ \\
$\lambda_C$ & 0.85 & 0.85 \\
$\sigma_{\mathrm{LJ}}$ (\AA) & 2.22 & 4.85 \\
$\varepsilon_{\mathrm{LJ}}$ (kJ mol$^{-1}$) & 1.472 & 0.077 \\
\hline
\end{tabular}
\end{table}

\subsection{Water force fields}

The TIP4P/05 water model \cite{Abascal2005} represents water as a rigid, non-polarizable molecule while preserving the molecular geometry of the original TIP4P model. The intermolecular interactions between water molecules are described by a combination of Lennard--Jones (LJ) and Coulombic potentials, as defined in Eq.~\ref{eq:potential}. As in the TIP4P family of water models, each hydrogen atom carries a positive partial charge, while the negative charge is assigned to a massless interaction site (M), located along the bisector of the H--O--H angle at a distance $l_{\mathrm{OM}}$ from the oxygen atom.

The molecular geometry and corresponding force-field parameters of the TIP4P/05 model are listed in Table~\ref{tip4p_parameters}. For the TIP4P/05 water model, the electrostatic scaling factors defined in Eq.~\ref{eq} are set to unity, i.e., $\lambda_{\mathrm{O}} = \lambda_{\mathrm{H}} = 1$. It is worth noting that the TIP4P/05 model was subsequently superseded by the TIP4P/$\epsilon$ model, which was developed to provide an improved description of the thermodynamic and structural properties of liquid water \cite{tip4pe}.

\begin{table}[ht] \centering \caption{Force-field parameters for the TIP4P/05 water model. The charge at the M site is given by $q_M=-2q_H$.} \label{tip4p_parameters} \begin{tabular}{|lccccccc|} \hline Model & $r_{\mathrm{OH}}$ / \AA & $\theta$ / $^\circ$ & $q_H$ / $e$ & $q_M$ / $e$ & $r_{\mathrm{OM}}$ / \AA & $\sigma$ / \AA & $(\epsilon/k_B)$ / K \\ \hline TIP4P/05 & 0.9572 & 104.52 & 0.5564 & $-1.1128$ & 0.1546 & 3.1589 & 93.2 \\ \hline \end{tabular} \end{table}

\section{ simulation details}
Molecular dynamics (MD) simulations were carried out with the GROMACS 2024
software package \cite{Abraham2024}. The equations of motion were propagated
using the leap-frog integration scheme \cite{Allen2017,Abraham2024}
with an integration step of $2$~fs. Electrostatic interactions beyond
the real-space cutoff were calculated using the Particle Mesh Ewald (PME)
approach \cite{Essmann1995}. A cutoff distance of $10$~\AA was employed for
the direct-space Coulomb interactions, whereas the reciprocal-space
contribution was calculated on a Fourier mesh with a grid spacing of
$1.2$~\AA using fourth-order spline interpolation. Periodic boundary
conditions were imposed along the three Cartesian directions, and a
cubic simulation cell was employed throughout the calculations.

The intramolecular geometry of the water molecules was maintained by means
of the LINCS constraint algorithm \cite{Hess1997}. The temperature was
regulated with a Nos\'e--Hoover thermostat, with $\tau_T=0.5$~ps
\cite{Tuckerman2001,Tuckerman2010}, while the pressure was controlled using
the Parrinello--Rahman barostat with a pressure-coupling time constant of
$\tau_P=1.0$~ps \cite{Tuckerman2010,Abraham2024}.

The simulation system was defined according to System 4 in the work of Manzanilla-Granados et al\cite{ManzanillaGranados2015}, figure \ref{10m}. In this system, a NaCl crystal is initially placed in the centre of a dilute aqueous NaCl solution. The initial solution has a molality of approximately 2 mol kg$^{-1}$ and is equilibrated for 20 ns. The subsequent calculations are performed over a 500 ns trajectory. The number of water molecules and NaCl ion pairs in the aqueous phase was varied according to the calculations and the selected $r_{\mathrm{cut}}$ values.

The molality, $m$, was determined from the number of NaCl ion pairs in the aqueous phase, $N_{\mathrm{NaCl}}$, excluding the ions belonging to the crystal, and the number of water molecules, $N_{\mathrm{H_2O}}$, according to

\begin{equation}
m =
\frac{N_{\mathrm{NaCl}}}
{N_{\mathrm{H_2O}}M_{\mathrm{H_2O}}}
\times 1000,
\end{equation}

where $M_{\mathrm{H_2O}}=18.015~\mathrm{g,mol^{-1}}$ is the molar mass of water. Alternatively, when the total number of ions in the aqueous phase, $N_{\mathrm{ions}}$, is used, the molality can be expressed as

\begin{equation}
m =
\frac{N_{\mathrm{ions}}}
{2N_{\mathrm{H_2O}}M_{\mathrm{H_2O}}}
\times 1000,
\end{equation}

where the factor of two accounts for the two ions comprising each NaCl formula unit.

\section{Results}

\subsection{Structural criterion for the identification of the crystalline phase}

The identificable of the crystalline NaCl phase is a critical aspect of the numerical determination of solubility. In the present approach, the radial distribution function, $g_{\mathrm{Na-Cl}}(r)$, is used to characterize the local structure of the NaCl crystal and to define a distance criterion for distinguishing ions associated with the crystalline phase from those belonging to the aqueous solution. The first maximum of $g_{\mathrm{Na-Cl}}(r)$ corresponds to the nearest-neighbor Na--Cl interactions, while the subsequent minimum separates the first coordination shell from the more distant structural correlations.

The conventional definition of the first coordination shell is obtained by integrating the radial distribution function up to its first minimum,

\begin{equation}
CN_{\mathrm{Na-Cl}}(r)
=
4\pi\rho_{\mathrm{Cl}}
\int_{0}^{r}
g_{\mathrm{Na-Cl}}(r')\,r'^2\,dr'
\label{eq:coordination_number}
\end{equation}

where $\rho_{\mathrm{Cl}}$ is the number density of chloride ions. The position of the first minimum therefore provides a physically meaningful distance for determining the first-shell coordination number. However, in the present work, the cutoff distance $r_{\mathrm{cut}}$ is not used exclusively to determine the first coordination shell. Instead, it is employed as a structural criterion to identify whether an ion belongs to the extended crystalline network.

This distinction is important because the crystalline NaCl phase is characterized by long-range structural connectivity that extends beyond the first coordination shell. Consequently, restricting the identification of crystalline ions strictly to the first minimum of $g_{\mathrm{Na-Cl}}(r)$ may exclude ions that remain structurally connected to the crystal lattice. Conversely, selecting a cutoff within the descending region of the first peak would truncate the first coordination shell and could lead to an incomplete representation of the crystalline phase. For this reason, the cutoff distance is selected according to the structural features of $g_{\mathrm{Na-Cl}}(r)$, with particular attention to the region separating the first and second coordination features.

In the present analysis, the first minimum of $g_{\mathrm{Na-Cl}}(r)$ is used to characterize the local coordination environment, whereas an extended cutoff associated with the onset of the second structural feature is employed to identify the crystalline phase. This criterion preserves the spatial connectivity of the ionic lattice while excluding ions that are clearly associated with the surrounding aqueous phase. To validate the choice of $r_{\mathrm{cut}}$, we performed a systematic analysis over a range of cutoff distances and compared the resulting structural properties with those reported by Benavides et al.~\cite{Benavides2017}. The regions, presented in Figure~\ref{g(r)mad}, was used to identify the cutoff that provides the best agreement with the previously reported results.

	
	\begin{figure}[H]
		\centering
		 {\includegraphics[clip,width=12cm,angle=0]{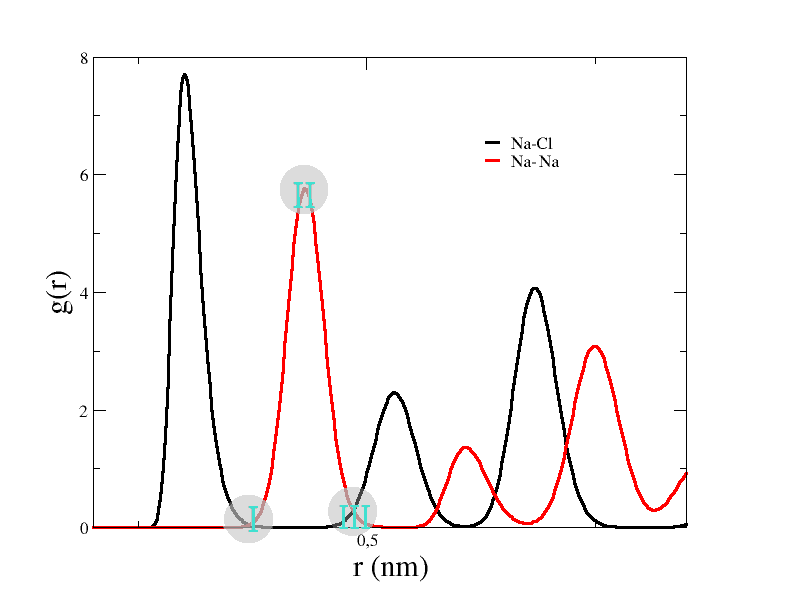}}
					
	 \caption{Radial distribution functions, $g(r)$, for the NaCl/Madrid model at 300 K and 1 bar. The figure shows the Na$^+$--Cl$^-$ and Na$^+$--Na$^+$ pair correlations, with the relevant regions used to determine the structural criteria for calculating the solubility of the NaCl--H$_2$O system indicated by appropriate labels.}
		\label{g(r)mad}
	\end{figure}

To assess the robustness of this structural criterion, the dependence of the calculated solubility on the selected cutoff distance is examined. Three values of $r_{\mathrm{cut}}$ are considered, including values within the descending region of the first peak, near the first minimum, and in the region preceding the second maximum of $g_{\mathrm{Na-Cl}}(r)$. For each cutoff, the number of ions assigned to the crystalline phase is monitored as a function of simulation time. The appropriate cutoff is then evaluated not only from the radial distribution function but also from the temporal stability of the identified crystalline phase and the resulting solubility.

The number of ions associated with the crystalline phase, $N_{\mathrm{crystal}}(t)$, provides a direct measure of the evolution of the solid phase. At the saturation condition, the crystal should remain stable on the time scale of the simulation, such that

\begin{equation}
\left\langle
\frac{dN_{\mathrm{crystal}}}{dt}
\right\rangle
\approx 0.
\label{eq}
\end{equation}

For an undersaturated solution, a systematic decrease in $N_{\mathrm{crystal}}(t)$ indicates net dissolution, whereas an increase in $N_{\mathrm{crystal}}(t)$ indicates net crystal growth and therefore a supersaturated solution. The saturation concentration is consequently determined from the concentration at which the crystalline phase remains statistically stable.

The use of an extended structural cutoff is therefore not intended to redefine the first coordination shell. Rather, it provides a criterion for identifying the connectivity of the crystalline phase during the dissolution process. The first minimum of $g_{\mathrm{Na-Cl}}(r)$ remains the appropriate structural reference for the local coordination analysis, while the extended cutoff is evaluated according to its ability to distinguish the crystalline and aqueous environments and to provide a stable and physically consistent determination of the solid--liquid equilibrium.

This procedure provides an internally consistent numerical criterion for the solubility calculation because the same structural definition is applied throughout the simulations. Importantly, the cutoff is determined from the molecular structure of the crystal and its sensitivity is explicitly evaluated rather than being selected solely to reproduce the experimental solubility. Consequently, agreement between the calculated and experimental solubilities constitutes an independent validation of the molecular model and of the proposed numerical procedure

\subsubsection{TIP4P/05 - NaCl/Madrid}

Benavides et al.~\cite{Benavides2017} reported a NaCl solubility of $5.7\pm0.2$~mol/kg for the TIP4P/05--NaCl/Madrid force-field combination, determined from the equality of the chemical potentials of NaCl in the solid and aqueous phases. This value is further supported by the independent results of Chattopadhyay et al.~\cite{Chattopadhyay2025}, who obtained essentially the same solubility using a Kirkwood--Buff (KB) theory-based approach. The agreement between these two independent methodologies is particularly relevant because the chemical-potential calculation and the Kirkwood--Buff approach rely on different theoretical frameworks for determining the solubility limit.

The consistency between these results provides a well-established reference value for assessing the solid--liquid behavior of the TIP4P/05--NaCl/Madrid model. In the present work, we therefore focus on this force-field combination and use the reported solubility of $5.7\pm0.2$~mol/kg as a reference for evaluating the criteria employed to identify the onset of the crystalline phase. In particular, the agreement with an independently determined solubility value provides a stringent benchmark for examining how the choice of the ion--ion cutoff and the resulting structural classification affect the estimated solubility. This approach allows the structural information obtained from the radial distribution functions and cluster analysis to be directly connected with a thermodynamically established solubility limit.

In the present work, we focus on numerically assessing and comparing this previously reported value using a direct coexistence approach. Specifically, a NaCl crystal is immersed in a dilute aqueous NaCl solution, following one of the four simulation setups proposed by Manzanilla-Granados et al.~\cite{ManzanillaGranados2015}. This configuration provides a direct numerical framework for evaluating the solubility and allows us to establish a quantitative comparison with the value reported by Benavides et al.~\cite{Benavides2017}.
	
	\begin{figure}[H]
		\centering
		 {\includegraphics[clip,width=12cm,angle=0]{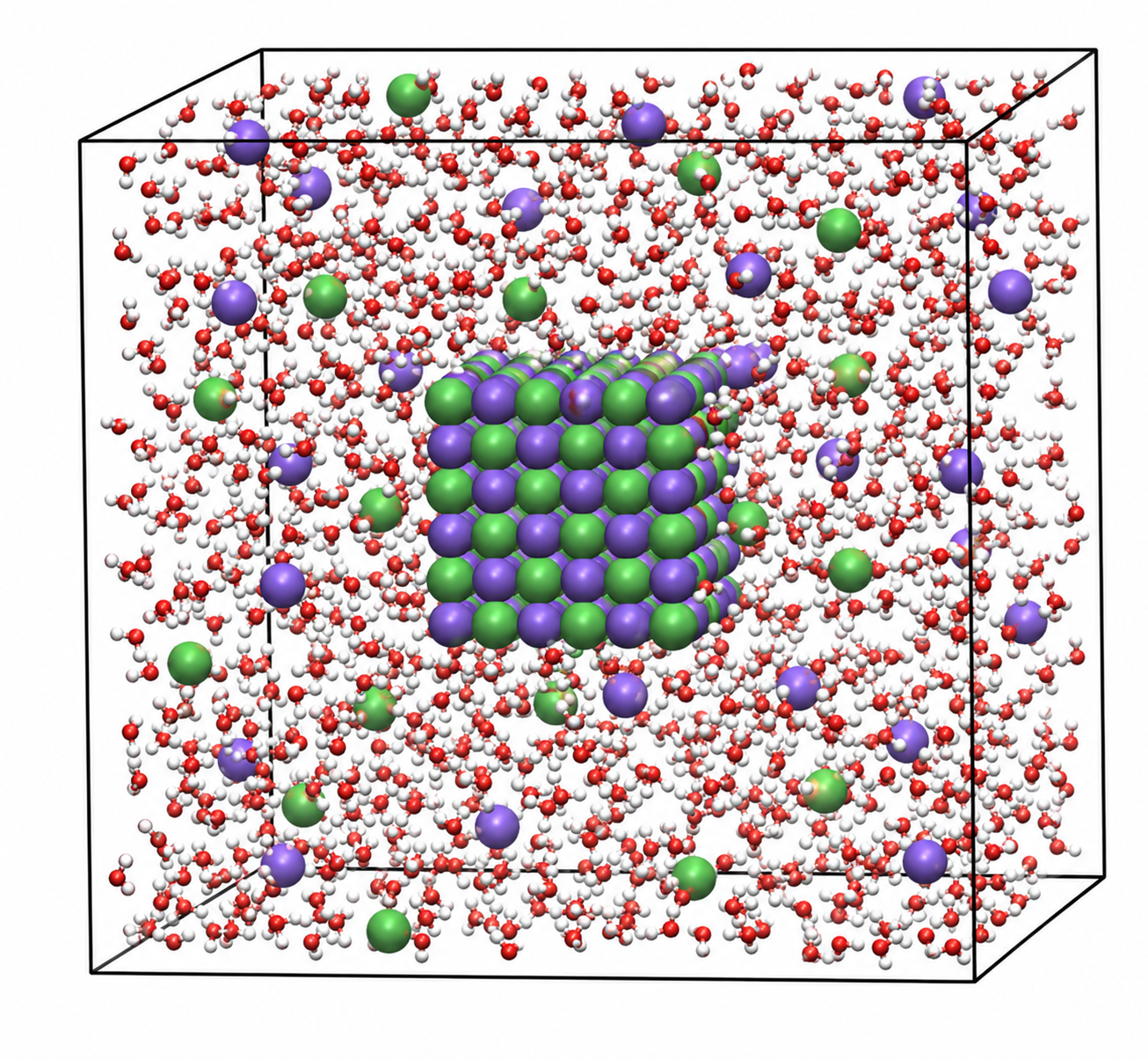}}
					
	 \caption{NaCl crystal immersed in a dilute aqueous NaCl solution: a model system for numerical solubility calculations.}
		\label{10m}
	\end{figure}
The initial configuration consists of a NaCl crystal immersed in an aqueous NaCl solution, as illustrated in Figure~\ref{10m}. During the simulation, the evolution of the crystal is monitored to determine whether dissolution or crystal growth occurs. A cutoff distance is defined from the radial distribution function, $g(r)$, of the crystalline system at 300~K and 1~bar. This cutoff is then used to distinguish ions belonging to the crystalline phase from those that remain dissolved in the aqueous solution. The solubility is subsequently calculated from the molality of the dissolved ions, considering only the ions that are not identified as part of the crystal and the surrounding water molecules.

This procedure is particularly sensitive to the choice of the cutoff distance, since this parameter determines which ions are classified as belonging to the crystalline phase and which are considered dissolved. Therefore, an appropriate structural criterion based on the characteristic features of the ionic $g(r)$ is essential for obtaining a reliable estimate of the solubility.

As shown in Figure~\ref{rcut}, three characteristic distances are selected as possible cutoff values based on the structural features of the radial distribution functions. These cutoff distances are subsequently used in the algorithm proposed by Sevik et al.~\cite{Sevick1988} to identify the ions belonging to the crystalline phase. Consequently, the calculated solubility exhibits a dependence on the selected cutoff distance, as summarized in Table~\ref{solubility_rcut}. This analysis allows us to assess the sensitivity of the calculated solubility to the structural criterion used to distinguish crystalline and dissolved ions.

\begin{table}[ht]
\centering
\caption{Dependence of the calculated solubility on the cutoff distance, $r_{\mathrm{cut}}$, used to identify the ions belonging to the crystalline phase.}
\label{solubility_rcut}
\begin{tabular}{|ccccccc|}
\hline
Region&$r_{\mathrm{cut}}$  & Maximum  & Total  & Ions in  & Water  & Molality  \\

&(nm)& cluster &ions&solution&molecules&$m$ (mol/kg)\\

\hline
III & 0.4848 & 481 & 500 & 19 & 1355 & 0.39 \\

& 0.4500 & 354 & 500 & 146 & 1355 & 2.99 \\
& 0.4400 & 242 & 500 & 258 & 1355 & 5.29 \\
II & 0.4323 & 158 & 500 & 342 & 1355 & 7.01 \\
& 0.4200 & 87 & 500 & 413 & 1355 & 8.47 \\

I & 0.3718 & 31 & 500 & 469 & 1355 & 9.61 \\
\hline
\end{tabular}
\end{table}

To assess the sensitivity of the calculated solubility to the structural criterion used to identify the solid phase, the dependence of the calculated solubility on the Na--Cl cutoff distance, $r_{\mathrm{cut}}$, was investigated. The resulting data show a pronounced and nonlinear decrease in the calculated solubility as $r_{\mathrm{cut}}$ is increased, indicating that the choice of cutoff distance is not merely a numerical detail but can directly influence the identification of the crystalline phase.

\begin{figure}[htbp]
    \centering

    \begin{subfigure}[b]{0.5\textwidth}
        \centering
        \includegraphics[width=\textwidth]{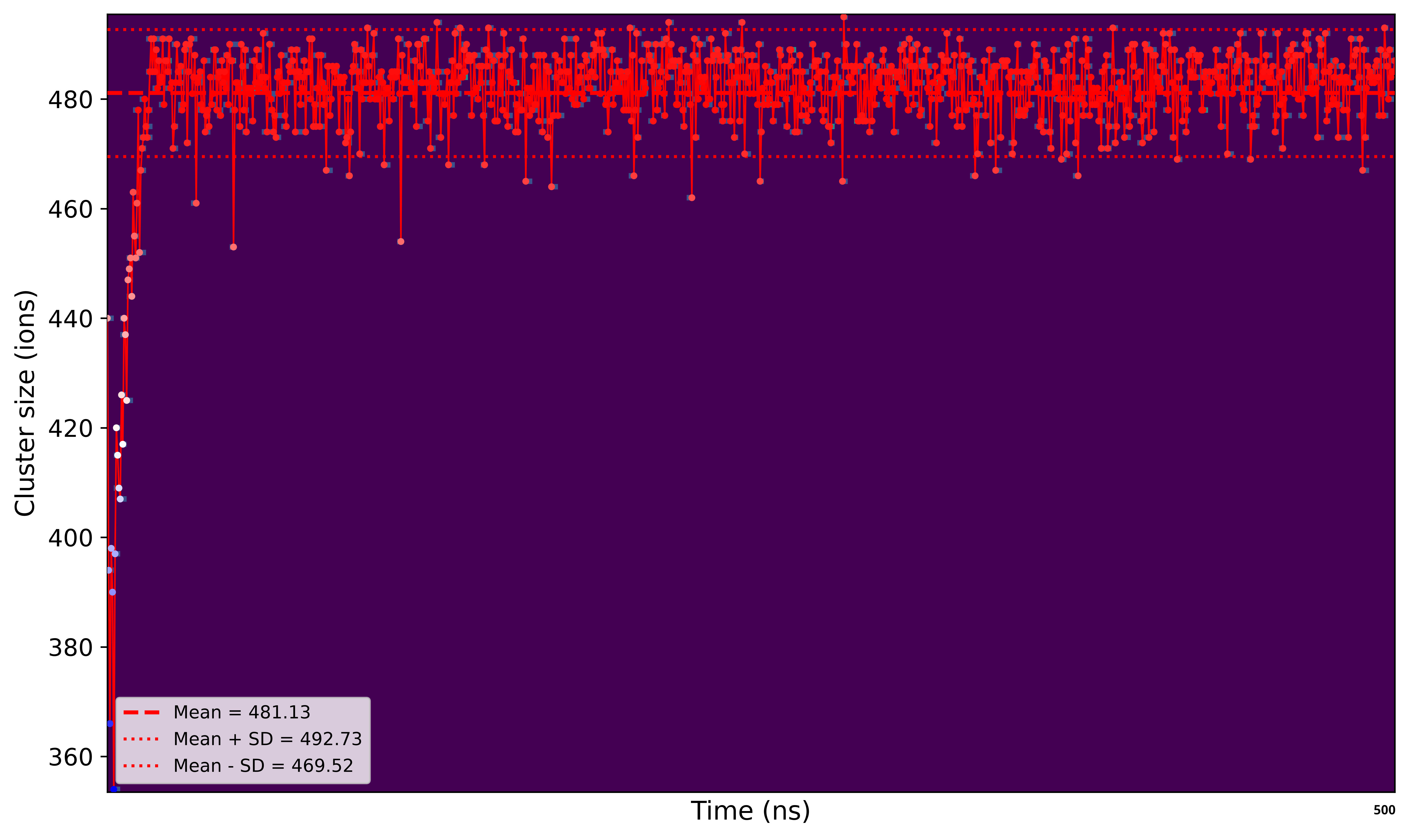}
        \caption{$r_{\mathrm{cut}}=0.4848$ nm}
        \label{fig:rcut04848}
    \end{subfigure}
    \hfill
    \begin{subfigure}[b]{0.5\textwidth}
        \centering
        \includegraphics[width=\textwidth]{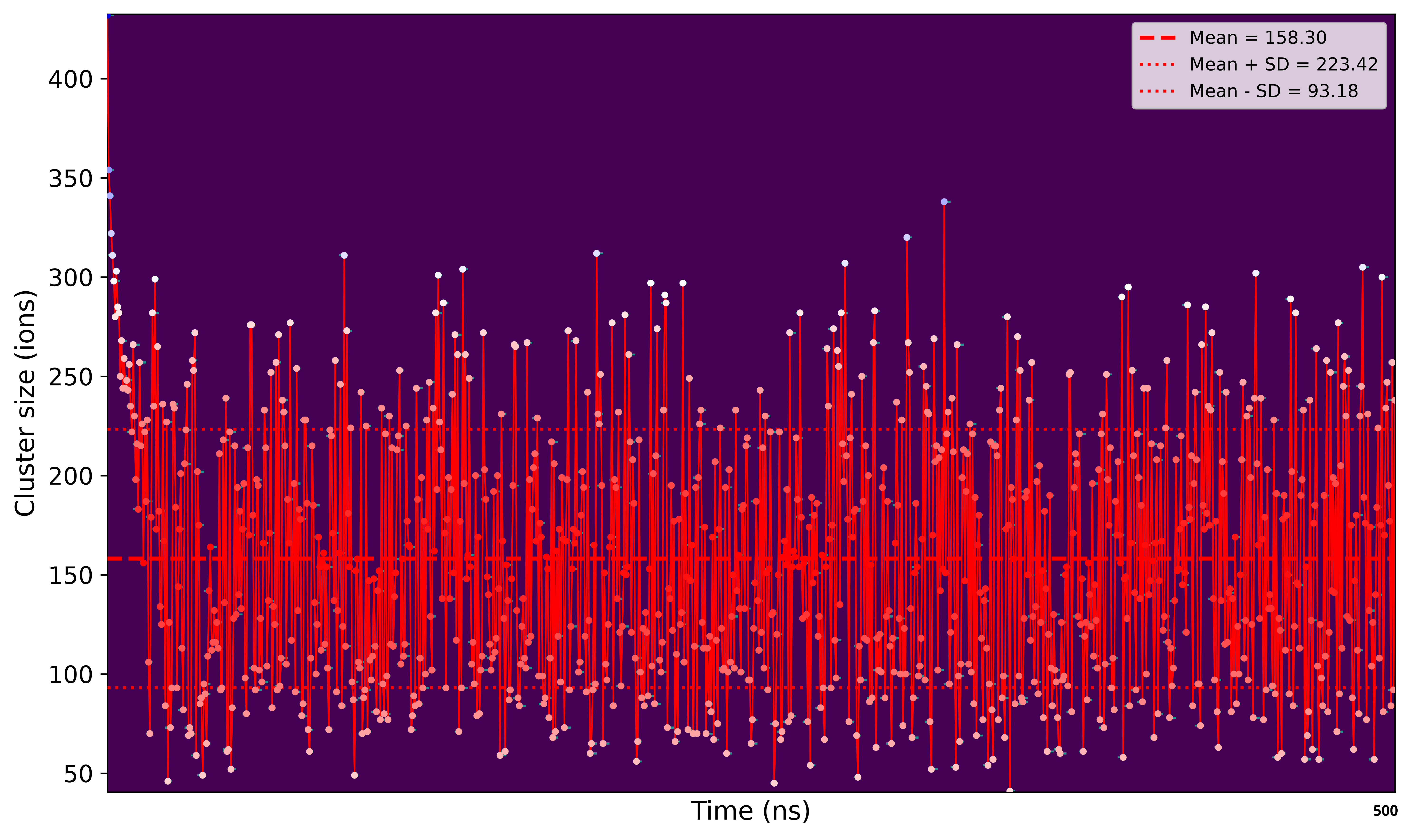}
        \caption{$r_{\mathrm{cut}}=0.4323$ nm}
        \label{fig:rcut04323}
    \end{subfigure}
    \hfill
    \begin{subfigure}[b]{0.5\textwidth}
        \centering
        \includegraphics[width=\textwidth]{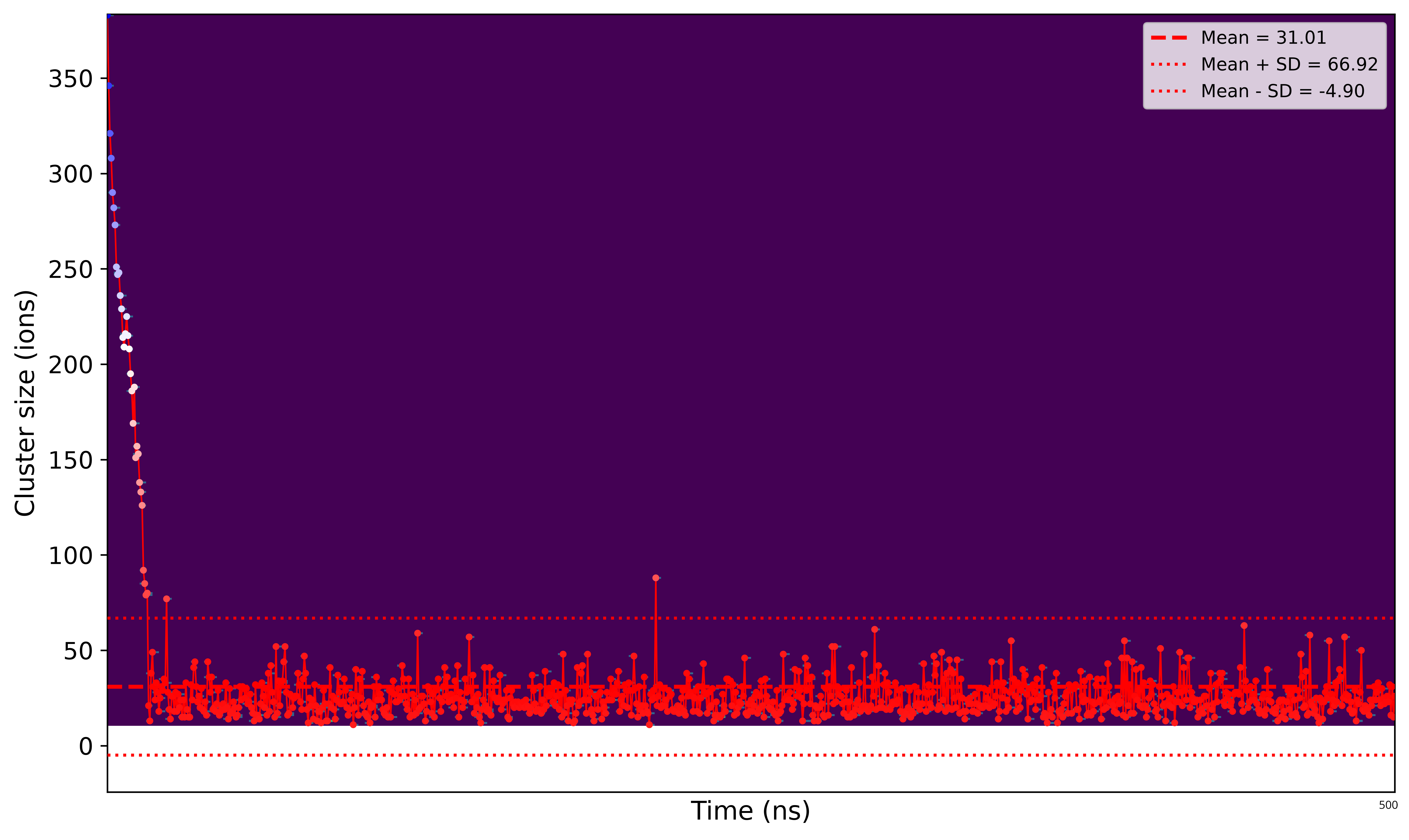}
        \caption{$r_{\mathrm{cut}}=0.3718$ nm}
        \label{fig:rcut03718}
    \end{subfigure}

    \caption{Time evolution of the crystal size for the three cutoff distances considered. 
    (a) $r_{\mathrm{cut}}=0.4848$ nm, (b) $r_{\mathrm{cut}}=0.4323$ nm, and 
    (c) $r_{\mathrm{cut}}=0.3718$ nm, for TIP4P/05 - NaCl/Madrid}
    \label{rcut}
\end{figure}

The temporal evolution of the crystal size obtained from the simulations is shown below. It is important to note that all simulations were performed over a total simulation time of $500~\mathrm{ns}$, using different values of the Na--Cl cutoff distance, $r_{\mathrm{cut}}$. In figure~\ref{fig:rcut04848}, a cutoff distance of $r_{\mathrm{cut}}=0.4848~\mathrm{nm}$ was employed, whereas figure~\ref{fig:rcut04323} corresponds to $r_{\mathrm{cut}}=0.4323~\mathrm{nm}$. Finally, figure~\ref{fig:rcut03718} shows the results obtained using $r_{\mathrm{cut}}=0.3718~\mathrm{nm}$. These three cutoff distances correspond to characteristic distances identified from the Na--Cl radial distribution function, $g_{\mathrm{Na-Cl}}(r)$, as shown in figure~\ref{g(r)mad}.

The results illustrate that the criterion used to define the Na--Cl connectivity has a direct effect on the identification and temporal evolution of the crystalline phase. In particular, changing $r_{\mathrm{cut}}$ modifies the number of Na--Cl contacts considered to belong to the same connected structure and, consequently, affects the apparent size of the crystal during the simulation.

	
	\begin{figure}[H]
		\centering
		 {\includegraphics[clip,width=12cm,angle=-90]{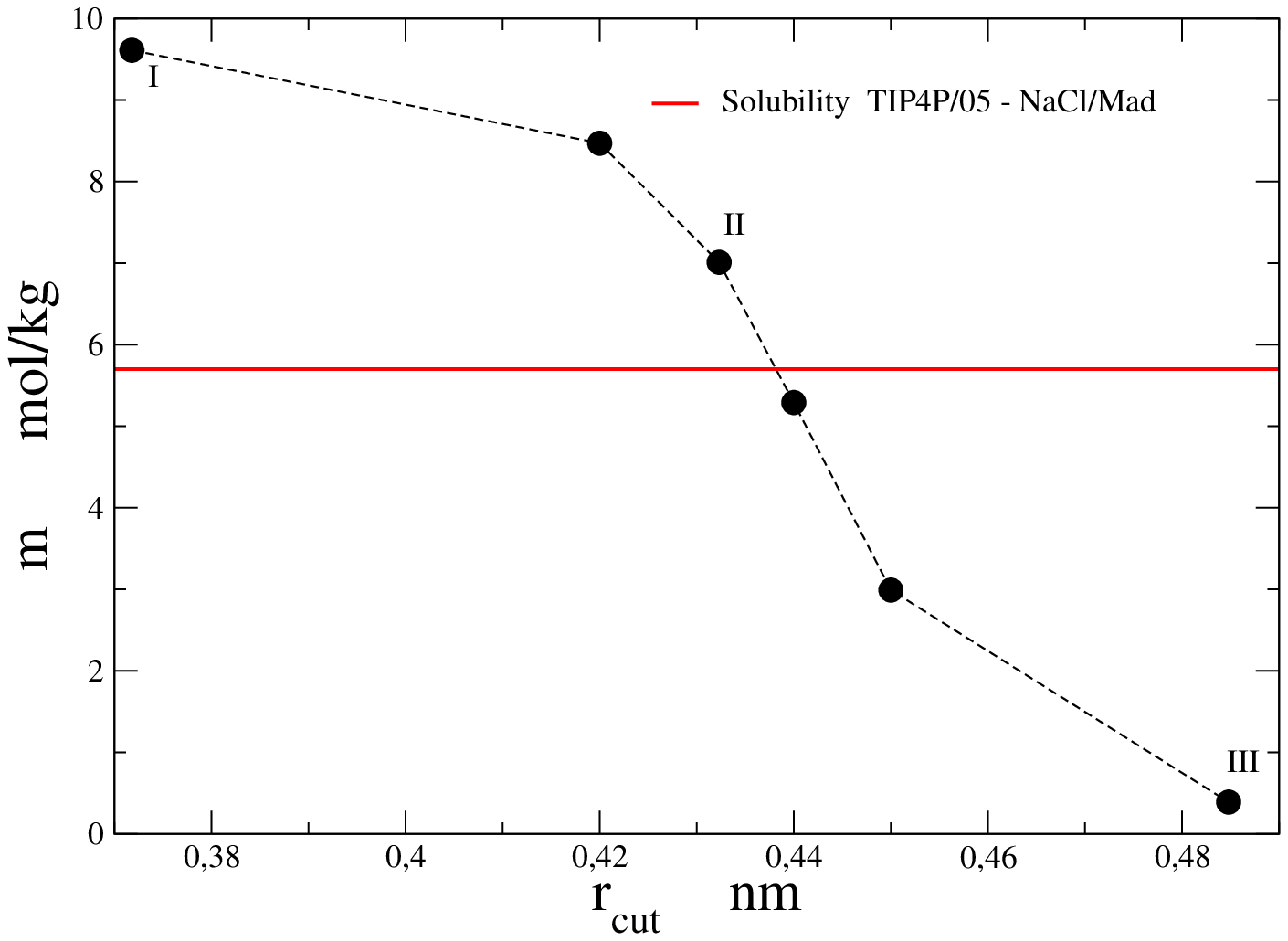}}
					
	 \caption{Dependence of the calculated NaCl solubility on the cutoff distance
$r_{\mathrm{cut}}$ for the NaCl/Madrid--TIP4P/05 model.}
		\label{solub-rcut}
	\end{figure}

Figure~\ref{solub-rcut} summarizes the dependence of the calculated solubility, $S$, on the cutoff distance, $r_{\mathrm{cut}}$. 

To quantitatively characterize the nonlinear dependence of the solubility on the concentration parameter, the calculated data were fitted using a third-order polynomial of the form

\begin{equation}
S(x)=ax^{3}+bx^{2}+cx+d,
\label{eq}
\end{equation}

where $S$ denotes the solubility and $x$ is the concentration parameter. A least-squares fit to the six calculated data points yielded

\begin{equation}
S(x)=31951.36298x^{3}
-41789.87840x^{2}
+18029.91917x
-2559.26290.
\label{eq}
\end{equation}

The inflection point of the fitted curve was determined from the condition that the second derivative vanishes,

\begin{equation}
\frac{d^{2}S}{dx^{2}}=6ax+2b=0,
\label{eq}
\end{equation}

which gives

\begin{equation}
x_{\mathrm{inf}}
=-\frac{b}{3a}
=0.435974.
\label{eq}
\end{equation}

Substitution of the inflection-point concentration into the cubic polynomial gives

\begin{equation}
S(x_{\mathrm{inf}})=5.88634.
\label{eq}
\end{equation}

The solubility predicted at the inflection point is consistent with the previously reported value of $5.7\pm0.2$. In fact, the reported uncertainty defines the interval $5.5\leq S\leq5.9$, and the value obtained from the present cubic fit, $S(x_{\mathrm{inf}})=5.88634$, lies within this interval. The difference between the calculated value and the reported central value is $0.18634$, which is smaller than the reported uncertainty of $0.2$.

Thus, the cubic fit identifies an inflection point at $x_{\mathrm{inf}}=0.435974$ that corresponds to a solubility of $5.88634$, in agreement with the reported solubility within its uncertainty. This result indicates that the concentration associated with the change in curvature of the calculated solubility relationship occurs in a region consistent with the previously reported solubility. Therefore, the inflection point obtained from the cubic representation provides a quantitative criterion for identifying the characteristic concentration associated with the solubility limit.

\section{Conclusions}

In this work, the solid--liquid behavior and aqueous solubility of NaCl were investigated using molecular dynamics simulations with the TIP4P/05--NaCl/Madrid force-field combination. Particular attention was given to the identification of a physically meaningful ion--ion cutoff distance, $r_{\mathrm{cut}}$, for defining the crystalline phase. Rather than selecting the cutoff arbitrarily, its effect on the calculated solubility was systematically investigated through a series of 500~ns simulations at 300~K and 1~bar.

The calculated solubility exhibits a strongly nonlinear dependence on $r_{\mathrm{cut}}$, decreasing from 9.61 to 0.39~mol/kg as the cutoff increases from 0.3718 to 0.4848~nm. To quantify this behavior, the calculated data were fitted to a third-order polynomial. The analysis identifies an inflection point at $r_{\mathrm{cut}}=0.435974$~nm, at which the fitted solubility is 5.88634~mol/kg. This value is consistent with the previously reported solubility of $5.7\pm0.2$~mol/kg, since it lies within the corresponding interval of 5.5--5.9~mol/kg. The difference from the reported central value is only 0.18634~mol/kg, which is smaller than the reported uncertainty.

The fact that the solubility obtained at the inflection point falls within the uncertainty of the previously reported value is particularly relevant because the concentration was not imposed as a fitting constraint. Instead, the characteristic cutoff was obtained solely from the nonlinear dependence of the calculated solubility on the structural criterion used to identify the crystalline phase. This result suggests that $r_{\mathrm{cut}}\simeq0.436$~nm provides a physically meaningful structural criterion for distinguishing crystalline and dissolved NaCl populations in the TIP4P/05--NaCl/Madrid model.

These findings are consistent with the conclusions of Espinosa et al.~\cite{Espinosa2017}, who demonstrated the importance of an adequate description of the crystalline phase and highlighted the influence of finite-size effects on the determination of NaCl solubility from molecular simulations. Their results emphasize that reliable characterization of the solid phase is essential when determining solid--liquid equilibrium properties.

The present results can also be considered in the context of the more recent work of Chattopadhyay et al.~\cite{Chattopadhyay2025}, who employed the Kirkwood--Buff approach to determine NaCl solubility and showed that the predicted solubility is strongly connected with the microscopic organization and ion pairing in solution. From this perspective, the present structural analysis provides a complementary approach in which the definition of the crystalline population is explicitly connected to the calculated solubility. The agreement with the independently reported solubility indicates that the structural information contained in the ion distribution and cluster population can be used to identify a characteristic length scale associated with the solid--liquid transition.

Overall, the results demonstrate that the ion--ion cutoff should not be regarded merely as a technical parameter for cluster analysis. Its systematic variation provides useful information about the structural definition of the crystalline phase and its relationship with the calculated solubility. For the TIP4P/05--NaCl/Madrid force-field combination, the present analysis identifies $r_{\mathrm{cut}}=0.435974$~nm as a characteristic structural cutoff, yielding a solubility of 5.88634~mol/kg, in agreement with the literature value of $5.7\pm0.2$~mol/kg within the reported uncertainty. This approach therefore provides a simple and reproducible structural criterion that complements thermodynamic methods such as chemical-potential calculations, direct coexistence, and Kirkwood--Buff theory for the study of solid--liquid equilibria in ionic systems.

\section{ Acknowledgements}
RFA gratefully acknowledges SECIHTI for financial support through an SNII
fellowship and the Universidad Autónoma Metropolitana (UAM) for providing
access to the Yotla cluster and the computational facilities used in this work.


\begin{thebibliography}{99}

\bibitem{Rogers2005}
P. J. Rogers and A. J. Beck,
``Calculation of the solubility of salts in water by molecular simulation,''
\textit{J. Phys. Chem. B} \textbf{109}, 1258--1264 (2005).

\bibitem{Rogers2011}
P. J. Rogers,
``Molecular simulation of electrolyte solutions and solubility,''
\textit{J. Chem. Phys.} \textbf{135}, 034506 (2011).

\bibitem{Palmer2016}
J. C. Palmer,
R. J. Spear, and P. G. Debenedetti,
``Solubility and crystallization of sodium chloride from molecular simulation,''
\textit{J. Chem. Phys.} \textbf{145}, 154501 (2016).

\bibitem{Moucka2011}
F. Mou\v{c}ka, M. L{\'i}sal, W. R. Smith, and I. Nezbeda,
``Molecular simulation of aqueous electrolyte solutions: The osmotic ensemble,''
\textit{J. Chem. Phys.} \textbf{135}, 154102 (2011).

\bibitem{Ladd1989}
A. J. C. Ladd and L. V. Woodcock,
``Triple-point coexistence properties of the Lennard-Jones system,''
\textit{Chem. Phys. Lett.} \textbf{164}, 191--195 (1989).

\bibitem{Davidchack2003}
R. L. Davidchack and B. B. Laird,
``Direct calculation of the hard-sphere crystal-melt interfacial free energy,''
\textit{Phys. Rev. Lett.} \textbf{85}, 4751--4754 (2000).

\bibitem{Davidchack2005}
R. L. Davidchack and B. B. Laird,
``Simulation of the hard-sphere crystal-melt interface,''
\textit{J. Chem. Phys.} \textbf{118}, 7651--7657 (2003).
\bibitem{Espinosa2016}
J.~R. Espinosa, C.~Vega, and E.~Sanz,
``The nucleation of NaCl in water: A molecular dynamics study,''
\textit{The Journal of Chemical Physics},
\textbf{144}, 034501 (2016).

\bibitem{Benavides2017}
A. L. Benavides, M. A. Portillo, V. C. Chamorro, J. R. Espinosa,
J. L. F. Abascal, and C. Vega,
``A potential model for sodium chloride solutions based on the
TIP4P/2005 water model,''
\textit{J. Chem. Phys.},
vol. 147, no. 10, p. 104501, 2017.
doi: 10.1063/1.5001190.

\bibitem{ManzanillaGranados2015}
H. M. Manzanilla-Granados, H. Saint-Mart{\'i}n, R. Fuentes-Azcatl, and J. Alejandre,
``Direct coexistence methods to determine the solubility of salts in water from numerical simulations. Test case NaCl,''
\textit{J. Phys. Chem. B} \textbf{119}, 8389--8396 (2015).

\bibitem{Espinosa2016}
J. R. Espinosa, J. M. Young, H. Jiang, D. Gupta, C. Vega, E. Sanz, P. G. Debenedetti, and A. Z. Panagiotopoulos,
``On the calculation of solubilities via direct coexistence simulations: Investigation of NaCl aqueous solutions and Lennard-Jones binary mixtures,''
\textit{J. Chem. Phys.} \textbf{145}, 154111 (2016).

\bibitem{Steinhardt1983}
P. J. Steinhardt, D. R. Nelson, and M. Ronchetti,
``Bond-orientational order in liquids and glasses,''
\textit{Phys. Rev. B} \textbf{28}, 784--805 (1983).

\bibitem{tenWolde1995}
P. R. ten Wolde, M. J. Ruiz-Montero, and D. Frenkel,
``Simulation of homogeneous crystal nucleation close to coexistence,''
\textit{Phys. Rev. Lett.} \textbf{75}, 2714--2717 (1995).

\bibitem{Allen2017}
M. P. Allen and D. J. Tildesley,
\textit{Computer Simulation of Liquids},
2nd ed. (Oxford University Press, Oxford, 2017).

\bibitem{Leach2001}
A. R. Leach,
\textit{Molecular Modelling: Principles and Applications},
2nd ed. (Prentice Hall, Harlow, 2001).

\bibitem{Abascal2005}
J. L. F. Abascal and C. Vega,
A general purpose model for the condensed phases of water: TIP4P/2005,
J. Chem. Phys. \textbf{123}, 234505 (2005).
doi:10.1063/1.2121687.


\bibitem{kbre}
R. Fuentes-Azcatl and M. C. Barbosa,
Potassium bromide, KBr/$\varepsilon$: New Force Field,
\textit{Physica A: Statistical Mechanics and its Applications}
\textbf{491}, 480--489 (2018).
doi:10.1016/j.physa.2017.09.081.

\bibitem{nacle}
R. Fuentes-Azcatl and M. C. Barbosa,
Sodium Chloride, NaCl/$\varepsilon$: New Force Field,
J. Phys. Chem. B \textbf{120}, 2460--2470 (2016).
doi:10.1021/acs.jpcb.5b12584.
\bibitem{HgCl2e}
R. Fuentes-Azcatl and M. A. Palomino-Ovando,
HgCl$_2$/$\varepsilon$: A New Force Field for Solid and Aqueous Solution:
A Molecular Dynamics Study,
Research Square, Preprint (Version 1), 11 August 2025.
doi:10.21203/rs.3.rs-7093271/v1.

\bibitem{LeontyevStuchebrukhov2009}
I. V. Leontyev and A. A. Stuchebrukhov,
``Electronic Continuum Model for Molecular Dynamics Simulations,''
\textit{J. Chem. Phys.},
vol. 130, no. 8, p. 085102, 2009.
doi: 10.1063/1.3060164.

\bibitem{LeontyevStuchebrukhov2010}
I. V. Leontyev and A. A. Stuchebrukhov,
``Electronic Continuum Model for Molecular Dynamics Simulations
of Biological Molecules,''
\textit{J. Chem. Theory Comput.},
vol. 6, no. 5, pp. 1498--1508, 2010.
doi: 10.1021/ct9005807.

\bibitem{tip4pe}
R. Fuentes-Azcatl and J. Alejandre,
Non-Polarizable Force Field of Water Based on the Dielectric Constant:
TIP4P/$\varepsilon$,
J. Phys. Chem. B \textbf{118}, 1263--1272 (2014).
doi:10.1021/jp410865y.
\bibitem{Abraham2024}
M. J. Abraham, T. Murtola, R. Schulz, S. P\aa{}ll,
J. C. Smith, B. Hess, and E. Lindahl,
GROMACS 2024.0,
Zenodo (2024).
doi:10.5281/zenodo.10589643.

\bibitem{Essmann1995}
U. Essmann, L. Perera, M. L. Berkowitz, T. Darden,
H. Lee, and L. G. Pedersen,
A smooth particle mesh Ewald method,
\textit{J. Chem. Phys.} \textbf{103}, 8577--8593 (1995).
doi:10.1063/1.470117.

\bibitem{Hess1997}
B. Hess, H. Bekker, H. J. C. Berendsen, and J. G. E. M. Fraaije,
LINCS: A linear constraint solver for molecular simulations,
\textit{J. Comput. Chem.} \textbf{18}, 1463--1472 (1997).
doi:10.1002/(SICI)1096-987X(199709)18:12<1463::AID-JCC4>3.0.CO;2-H.

\bibitem{Tuckerman2010}
M. E. Tuckerman,
\textit{Statistical Mechanics: Theory and Molecular Simulation},
Oxford University Press, Oxford (2010).

\bibitem{Tuckerman2001}
M. E. Tuckerman, Y. Liu, G. Ciccotti, and G. J. Martyna,
Non-Hamiltonian molecular dynamics: Generalizing Hamiltonian phase
space principles to non-Hamiltonian systems,
\textit{J. Chem. Phys.} \textbf{115}, 1678--1702 (2001).
doi:10.1063/1.1383046.

\bibitem{Chattopadhyay2025}
A.~Chattopadhyay, V.~Mandalaparthy, and N.~F.~A.~van~der~Vegt,
``Determination of aqueous solubility of NaCl in molecular dynamics simulation using the Kirkwood--Buff method,''
\textit{J. Chem. Phys.} \textbf{162}, 174116 (2025).
doi:10.1063/5.0264104.

\bibitem{Sevick1988}
E. M. Sevick, P. A. Monson, and J. M. Ottino,
Monte Carlo calculations of cluster statistics in continuum models
of composite morphology,
\textit{J. Chem. Phys.} \textbf{88}, 1198--1206 (1988).
doi:10.1063/1.454674.


\end{thebibliography}
\end{document}